\documentclass[twocolumn]{aastex631}

\usepackage{amsmath}
\shorttitle{C~I traces the disk surface in IM~Lup}
\shortauthors{Law et al.}
\graphicspath{{./}{figures/}}
\begin{document}

\title{\ion{C}{1} Traces the Disk Atmosphere in the IM~Lup Protoplanetary Disk}

\author[0000-0003-1413-1776]{Charles J. Law}
\altaffiliation{NASA Hubble Fellowship Program Sagan Fellow}
\affiliation{Department of Astronomy, University of Virginia, Charlottesville, VA 22904, USA}

\author[0000-0002-2692-7862]{Felipe Alarc\'on}
\affiliation{Department of Astronomy, University of Michigan, 323 West Hall, 1085 South University Avenue, Ann Arbor, MI 48109, USA}

\author[0000-0003-2076-8001]{L. Ilsedore Cleeves}
\affiliation{Department of Astronomy, University of Virginia, Charlottesville, VA 22904, USA}

\author[0000-0001-8798-1347]{Karin I. \"Oberg}
\affiliation{Center for Astrophysics \textbar\, Harvard \& Smithsonian, 60 Garden St., Cambridge, MA 02138, USA}

\author[0000-0002-4044-8016]{Teresa Paneque-Carre\~{n}o}
\affiliation{European Southern Observatory, Karl-Shwarzschild-Strasse 2, D-85748 Garching bei Munchen, Germany}



\begin{abstract}
The central star and its energetic radiation fields play a vital role in setting the vertical and radial chemical structure of planet-forming disks. We present observations that, for the first time, clearly reveal the UV-irradiated surface of a protoplanetary disk. Specifically, we spatially resolve the atomic-to-molecular~(\ion{C}{1}-to-CO) transition in the IM~Lup disk with ALMA archival observations of [\ion{C}{1}]~$^3$P$_1$-$^3$P$_0$. We derive a \ion{C}{1} emitting height of z/r~$\gtrsim$~0.5 with emission detected out to a radius of ${\approx}$600~au. Compared to other systems with \ion{C}{1} heights inferred from unresolved observations or models, the \ion{C}{1} layer in the IM~Lup disk is at scale heights almost double that of other disks, confirming its highly flared nature. \ion{C}{1} arises from a narrow, optically-thin layer that is substantially more elevated than that of $^{12}$CO~(z/r~$\approx$~0.3-0.4), which allows us to directly constrain the physical gas conditions across the \ion{C}{1}-to-CO transition zone. We also compute a radially-resolved \ion{C}{1} column density profile and find a disk-averaged \ion{C}{1}~column density of 2$\times10^{16}$~cm$^{-2}$, which is ${\approx}$3-20$\times$~lower than that of other disks with spatially-resolved \ion{C}{1}~detections. We do not find evidence for vertical substructures or spatially-localized deviations in \ion{C}{1} due, e.g., to either an embedded giant planet or a photoevaporative wind that have been proposed in the IM~Lup disk, but emphasize that deeper observations are required for robust constraints.
\end{abstract}
\section{Introduction} \label{sec:intro}

Protoplanetary disks show a high degree of vertical stratification due to both physical and chemical gradients \citep[e.g.,][]{Oberg23_REV} as well as a variety of dynamical processes \citep[e.g.,][]{Pinte22_PP}. The relative vertical distribution of material sets the overall disk structure and greatly influences what material is available to forming planets. The high~spatial resolution and sensitivity of the Atacama Large Millimeter/submillimeter Array (ALMA) has enabled a detailed mapping of the vertical emitting heights of numerous molecular species in disks \citep[e.g.,][]{pinte18, Law21, Paneque_MAPS}, but far fewer studies have focused on the atomic gas expected to be present in the disk atmosphere.

In particular, neutral atomic carbon (\ion{C}{1}) is thought to occupy a thin vertical region between the CO photodissociation and carbon ionization fronts, and is, in turn, bounded above by emission from (photo)ionized carbon \citep{Tielens85, Jonkheid04, vanDishoeck06}. This photodissociation region (PDR)-like emission structure reflects changing UV field strengths as shielding becomes more efficient in deeper disk layers. Here, we focus on \ion{C}{1}, since it is readily observable via its sub-millimeter forbidden lines and is expected to be the  dominant gas-phase carbon carrier at the disk heights where it emits.

\ion{C}{1} was originally detected in the DM~Tau protoplanetary disk by \citet{Tsukagoshi15}, followed by several subsequent disk detections \citep[e.g.,][]{Kama16_TWHya_HD100546, Sturm22, Pascucci23}. However, the majority of existing \ion{C}{1} detections are spatially unresolved. Even for those few existing spatially-resolved observations, most are at modest angular resolutions (${\gtrsim}0\farcs5$--1$^{\prime \prime}$) \citep[e.g.,][]{Alarcon22, Temmink23, Booth23_HD169142}. This has prohibited a detailed understanding of the \ion{C}{1} emission structure, namely, while it is widely assumed that \ion{C}{1} traces the disk upper layers, no observations to date have directly confirmed this or pinpointed the exact location of the transition region between \ion{C}{1} and the CO molecular layer. 

A detailed understanding of the \ion{C}{1} emitting region in disks is also critical for properly identifying and interpreting potential wind signatures or flows driven by embedded protoplanets. For instance, \citet{Alarcon22} discovered a localized \ion{C}{1} kinematic deviation thought to trace an embedded protoplanet in the HD~163296 disk. However, these conclusions remain tentative due to the limited angular resolution of the observations and the inability to accurately deproject the \ion{C}{1} rotation map due to a lack of information on the vertical \ion{C}{1} emitting surface. Additionally, the identification of photoevaporative winds remains elusive despite the potentially outsized role they play in shaping disk evolution and planet formation outcomes \citep[e.g.,][]{Hollenbach00, Matt05}. \ion{C}{1} may be a tracer of such winds but if present, their detection would require sensitive, spatially-resolved observations and a precise knowledge of the vertical \ion{C}{1} distribution. 

In this Letter, we present high-angular-resolution ALMA archival observations of the [\ion{C}{1}] $^3$P$_1$-$^3$P$_0$ line in the IM~Lup disk. Using these data, we extract the \ion{C}{1} emitting surface in a planet-forming disk for the first time and provide direct observational confirmation that \ion{C}{1} emission originates from the disk atmosphere. In Section \ref{sec:IMLup_disk}, we briefly describe the IM~Lup disk and summarize the ALMA observations in Section \ref{sec:observations_overview}. We present the \ion{C}{1} emission surfaces in Section \ref{sec:results}, along with radially-resolved \ion{C}{1} column densities. In Section \ref{sec:discussion}, we discuss the \ion{C}{1} vertical emission morphology in the context of the IM~Lup disk and thermochemical models. We summarize our conclusions in Section \ref{sec:conlcusions}.

\section{The IM~Lup Disk}  \label{sec:IMLup_disk}

IM~Lup is a young (${\sim}$1~Myr), approximately solar-mass, T~Tauri star located in the Lupus star-forming region \citep{Mawet12, Alcala17, Teague21} at a distance of $d=158$~pc \citep{Gaia18} that hosts a massive \citep{Zhang21, Lodato23} and unusually large protoplanetary disk extending to several hundreds of au in its millimeter dust \citep{Huang18_Elias,Andrews18}, NIR/scattered light \citep{Avenhaus18}, and molecular line emission \citep[][]{Panic09, Cleeves16_IMLup, LawMAPSIII}. The outer region of its $^{12}$CO gas disk, which reaches a maximum radius of ${\sim}$1000~au, is diffuse and envelope-like, potentially indicating the presence of an external photoevaporative wind \citep{Haworth17}. The IM~Lup disk is vertically-extended and highly-flared in its micron-sized dust distribution \citep{Avenhaus18, Rich21} and $^{12}$CO emission surface \citep{Huang17, pinte18, Law21}. There is also indirect kinematic evidence for at least one giant planet embedded within the IM~Lup disk \citep{Speedie22, Verrios22, Izquierdo23_DM2}.

The inclined nature of the IM~Lup disk coupled with its large radial and vertical extent make it an ideal source to locate the \ion{C}{1} emitting region. Since IM~Lup has been extensively observed at sub-mm wavelengths and has well-constrained models of its physical and chemical disk structure \citep[e.g.,][]{Cleeves16_IMLup, Zhang21}, we can compare the location of \ion{C}{1} with that of other disk structure tracers and infer the physical conditions at the location of the atomic carbon-to-molecular CO transition zone.

For all subsequent analysis, we adopt the following parameters for the IM~Lup system: PA=144.$^{\circ}$5, incl=47.$^{\circ}$5, v$_{\rm{sys}}$=4.5~km~s$^{-1}$, and M$_*$=1.1~M$_{\odot}$ \citep{Oberg21_MAPSI, Teague21}. 

\begin{figure*}
\centering
\includegraphics[width=\linewidth]{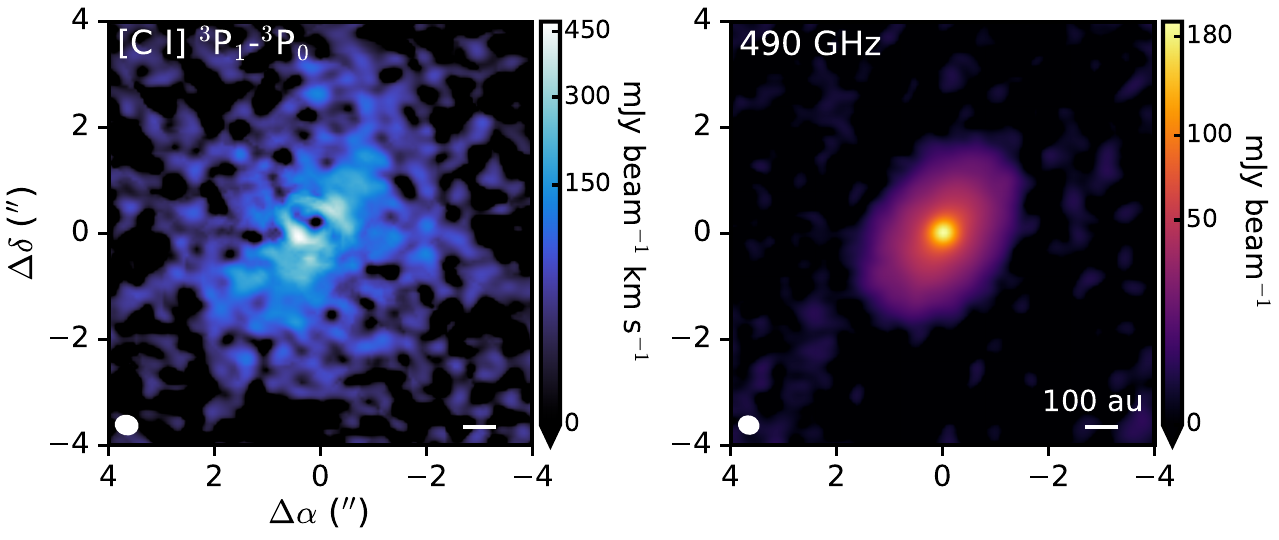}
\caption{[\ion{C}{1}] 1--0 zeroth moment map (left) and 490~GHz continuum image (right) of the IM~Lup disk. The synthesized beam and a scale bar indicating 100~au is shown in the lower left and right corner, respectively, of each panel.} 
\label{fig:figure_moments}
\end{figure*}

\section{Observations and Analysis}
\label{sec:observations_overview}

\subsection{Archival Data and Observational Details}
\label{sec:archival_data}

We made use of the ALMA archival project 2015.1.01137.S (PI: T. Tsukagoshi), which included Band~8 observations of the IM~Lup disk that were originally published in \citet{Pascucci23}. The IM~Lup disk was observed on 25 March 2016 for 9~min using 41 antennas with a projected baseline range of 15-443~m. The [\ion{C}{1}] $^3$P$_1$-$^3$P$_0$ (hereafter, [\ion{C}{1}] 1--0) line at 492.161~GHz was centered in a spectral window with a frequency resolution of 244~kHz (${\approx}$0.15~km~s$^{-1}$). The maximum recoverable scale (MRS) of the data is ${\approx}$2\farcs8, which is comparable to the overall radial extent of the [\ion{C}{1}] 1--0 emission. This limited MRS combined with the modest signal-to-noise ratio (SNR) of the observations likely results in the ``patchy'' appearance of the overall [\ion{C}{1}] 1--0 emission morphology evident in the line emission cubes (see Appendix \ref{sec:appendix_channel_maps}). Further details about these observations can be found in \citet{Pascucci23}.

\subsection{Self Calibration and Imaging}
\label{sec:selfcal_imaging}

The archival data was initially calibrated by ALMA staff using the ALMA calibration pipeline and the required version of CASA \citep{McMullin_etal_2007}, before switching to CASA \texttt{v5.4.0} for self calibration. We followed standard self-calibration strategies \citep[e.g.,][]{Oberg21_MAPSI}. We performed one round of phase self-calibration (\texttt{solint}=`inf') using the continuum and found similar improvement in the peak continuum SNR as reported by \citet{Pascucci23}. We then applied the self-calibration solutions to the full line image cubes and subtracted the continuum using the \texttt{uvcontsub} task with a first-order polynomial. 

We then switched to CASA \texttt{v6.3.0} for imaging. We used the \texttt{tclean} task to image [\ion{C}{1}] 1--0 with a Briggs weighting of \texttt{robust}$=$2 to maximize the SNR. We used a Keplerian mask generated with the \texttt{keplerian\_mask} \citep{rich_teague_2020_4321137} code that was based on the stellar and disk parameters of IM~Lup. The mask was visually inspected to ensure that it contained all emission present in the channel maps. Images were generated with channel spacings of 0.15~km~s$^{-1}$. This fine velocity resolution is critical to have the maximum number of channels in which the emission surface is clearly visible. All images were made using the ‘multi-scale’ deconvolver with pixel scales of [0,5,15,25] and were CLEANed down to a 4$\sigma$ level, where $\sigma=65$~mJy~beam$^{-1}$ was the RMS measured in a line-free channel of the dirty image. The resulting image had a beam size of 0\farcs41$\times$0\farcs35 and PA=76.$^{\circ}$9. We made a 490~GHz continuum image using the full bandwidth of the observations after flagging channels containing line emission with \texttt{robust}$=$0.5 and an elliptical mask. Otherwise, the CLEANing parameters were the same. We measured a continuum rms of $\sigma$=0.68~mJy and the resulting image had a beam size of 0\farcs37$\times$0\farcs32 and PA=73.$^{\circ}$4. To ensure that the [\ion{C}{1}] 1--0 images are properly centered, which is required for proper deprojection and derivation of accurate emitting heights, we used the \texttt{imfit} task to fit a 2D Gaussian to the continuum image. We found an offset of $\Delta \rm{R.A.}=+0\farcs75$ and $\Delta \rm{decl.}=-0\farcs02$, which is used to center the line images.

We generated a zeroth moment map of the [\ion{C}{1}] 1--0 line emission using \texttt{bettermoments} \citep{Teague18_bettermoments} with no flux threshold for pixel inclusion and the same Keplerian mask used for CLEANing. Figure \ref{fig:figure_moments} shows the [\ion{C}{1}] 1--0 zeroth moment map and 490~GHz continuum images. The full [\ion{C}{1}] 1--0 channel maps are presented in Appendix \ref{sec:appendix_channel_maps}.

\begin{figure*}
\centering
\includegraphics[width=1\linewidth]{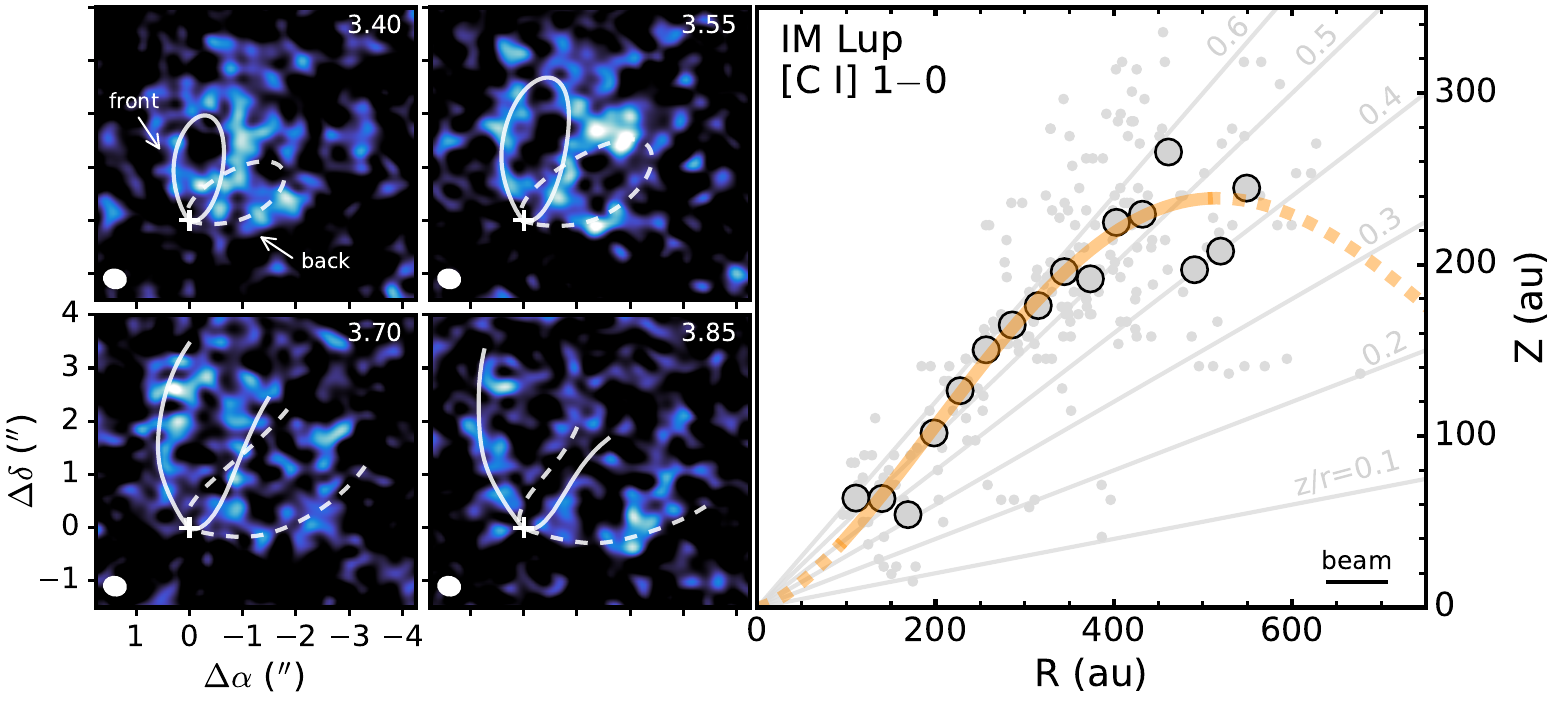}  
\vspace{0.1cm} \\
\includegraphics[width=0.9125\linewidth]{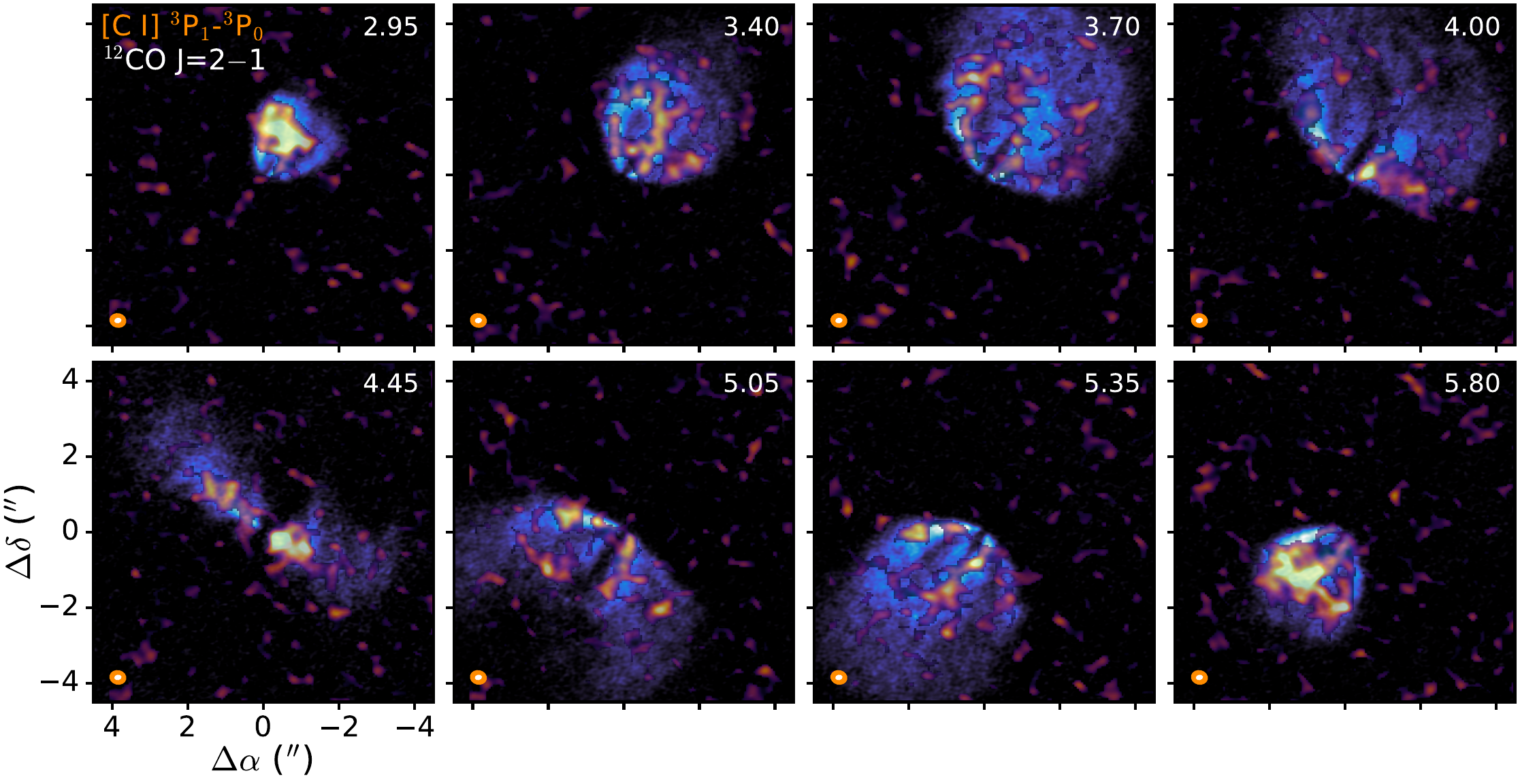}\hspace{38pt}
\caption{\textit{Top left:} Representative [\ion{C}{1}] 1--0 channels showing the upper (solid) and lower (dashed) emission surfaces with isovelocity contours of the best-fitting model overlaid. The LSRK velocity and synthesized beam are shown in the the upper right and bottom left corner, respectively, of each panel. \textit{Top right:} Derived [\ion{C}{1}] 1--0 emission surface showing both radially-binned (large points) and individual measurements (small points). The orange line shows the exponentially-tapered power law fit, where the solid line shows the radial range used in the fitting and the dashed line is an extrapolation. Lines of constant z/r from 0.1 to 0.6 are shown in gray. \textit{Bottom:} Channel maps of $^{12}$CO J=2--1 with [\ion{C}{1}] 1--0 emission overlaid in orange. The $^{12}$CO data are taken from \citet{Oberg21_MAPSI}. The synthesized beams are shown in the lower left corner of each panel.}
\label{fig:surface_extracted}
\end{figure*}

\subsection{Emission Surface Extraction}\label{sec:methods_sub_surfextr}

We derived vertical emission heights from the line emission image cubes using the \texttt{alfahor}\footnote{\url{https://github.com/teresapaz/alfahor}} Python code \citep{Paneque_MAPS}, which is based on the methodology originally presented in \citet{pinte18}. This approach requires hand-drawn masks for each channel where the emission surface can be visually distinguished. This approach is particularly useful due to the ``patchy" emission surfaces, which are relatively easy to identify by-eye but were found to confuse fully-automated retrieval techniques, e.g., \texttt{disksurf} \citep{disksurf_Teague}. While only the upper surface of the disk is used to define the extracted emission surfaces, the lower surface is often also clearly distinguished in the channel maps (see Figure \ref{fig:surface_extracted}). All extracted surfaces were then binned in two separate ways: (1) into radially bins equal to the FWHM of the beam major axis; (2) and by computing a moving average with a window size of the beam major axis FWHM.

We then fitted a parametric model to the [\ion{C}{1}] 1--0 emission surface in the form of an exponentially-tapered power law, which captures the inner flared surface and the plateau and turnover regions at larger radii \citep[e.g.,][]{Teague_19Natur, Law21}:

\begin{equation} \label{eqn:exp_taper}
z(r) = z_0 \times \left( \frac{r}{1^{\prime \prime}} \right)^{\phi} \times \exp \left(- \left[ \frac{r}{r_{\rm{taper}}} \right]^{\psi} \right)
\end{equation}

We used the Monte Carlo Markov Chain (MCMC) sampler implemented in \texttt{emcee} \citep{Foreman_Mackey13} to obtain the posterior probability distributions for $z_0$, $\phi$, $r_{\rm{taper}}$, and $\psi$. Each ensemble consisted of 64 walkers with 1000 burn-in steps and an additional 1000 steps to sample the posterior distribution function. We take the median values of the posterior distribution as the best-fitting values with the uncertainties given by the 16th and 84th percentiles. We found the following values: $z_0=0\farcs515^{+0.216}_{-0.129}$; $\phi=1.476^{+0.487}_{-0.398}$; $r_{\rm{taper}}=3\farcs917^{+1.373}_{-1.270}$; and $\psi=2.200^{+2.464}_{-1.490}$. The relatively higher uncertainties of $r_{\rm{taper}}$ and $\psi$ reflect the increased difficulty in obtaining emitting heights in the outer disk due to the lower SNR.

\section{Results} \label{sec:results}

\subsection{\texorpdfstring{[\ion{C}{1}]}{[C I]} 1--0 Emission Heights} \label{sec:CI_emitting_heights}

Figure \ref{fig:surface_extracted} shows the extracted [\ion{C}{1}] 1--0 emitting heights in the IM~Lup disk. The [\ion{C}{1}] 1--0 emission rises steeply ($z/r\gtrsim$~0.5) from 100~au to ${\approx}$400~au, beyond which it shows the typical flattening and turnover seen in other disk emission surfaces \citep[e.g.,][]{Teague_19Natur}, likely due to decreasing gas densities. Emitting heights interior to ${\approx}$100~au could not be derived due to the combination of our beam size and the central emission cavity (Figure \ref{fig:figure_moments}), while extraction of heights beyond 600~au was primarily limited by the lower SNRs at these larger radii. The [\ion{C}{1}] 1--0 emission is consistent with a Keplerian rotation pattern and we do not identify any significant non-Keplerian or large-scale emission components \citep[e.g.,][]{Sturm22}. To further demonstrate this, Figure \ref{fig:surface_extracted} also shows a representative set of channels maps with [\ion{C}{1}] 1--0 emission overlaid on that of $^{12}$CO J=2--1. The [\ion{C}{1}] 1--0 traces the same Keplerian rotation as $^{12}$CO but is clearly arising from higher disk elevations.

\subsection{\texorpdfstring{\ion{C}{1}}{C I} Emission Morphology and Column Density} \label{sec:Ncol}

To better understand the [\ion{C}{1}] 1--0 morphology emission, we generated an azimuthally-averaged line intensity radial profile. We used the radial profile function in the \texttt{GoFish} python package \citep{Teague19JOSS} to deproject the zeroth moment map (Figure \ref{fig:figure_moments}) along the derived [\ion{C}{1}] 1--0 emission surfaces (Figure \ref{fig:surface_extracted}). For emission originating from sufficiently elevated layers, it is necessary to consider its emitting surface during the deprojection process to compute accurate radial positions.

\begin{figure}
\centering
\includegraphics[width=\linewidth]{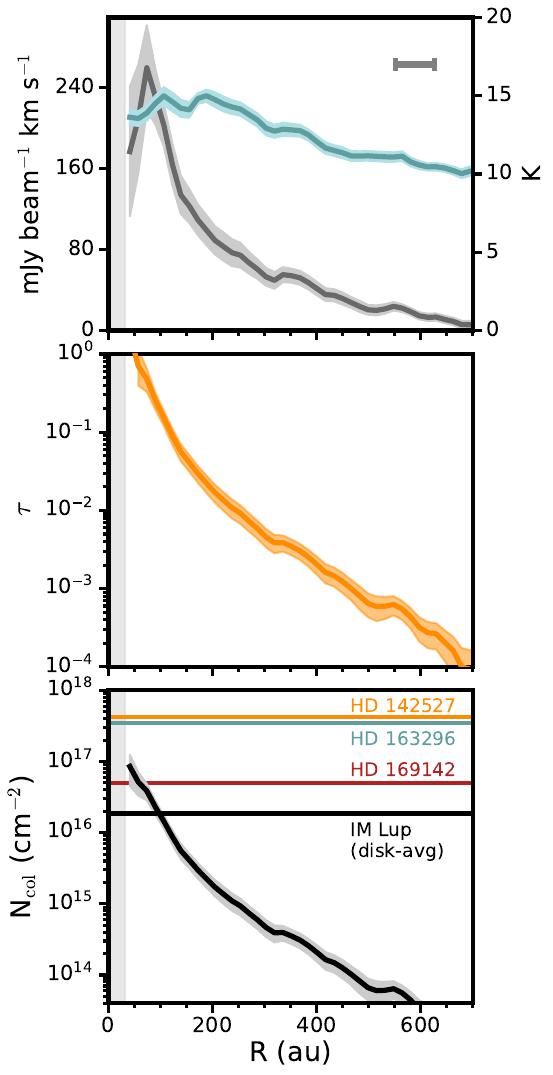}
\caption{\textit{Top}: Azimuthally-averaged [\ion{C}{1}] 1--0 radial intensity (gray) and peak intensity (blue) profiles deprojected using the derived emission surface. Shaded regions show the 1$\sigma$ uncertainty. The FWHM of the major axis of the synthesized beam is shown by a horizontal bar in the upper right. \textit{Middle}: Optical depth profile of [\ion{C}{1}] 1--0. \textit{Bottom}: Radial \ion{C}{1} column density profile. Both optical depth and N$_{\rm{col}}$ profiles assume thermal line widths and a constant excitation temperature of 50~K. Horizontal lines show values derived from spatially-resolved ALMA observations of HD~163296 \citep{Alarcon22}, HD~169142 \citep{Booth23_HD169142}, and HD~142527 \citep{Temmink23}. The horizontal shaded region indicates 1/2$\times$ the beam major axis in all panels.}
\label{fig:radial_profiles}
\end{figure}

The top panel of Figure \ref{fig:radial_profiles} shows the resulting radial profile. The [\ion{C}{1}] 1--0 emission takes the form of a central cavity, broad ring at ${\approx}$75~au, and extended emission out to a radius of ${\sim}$700~au. It is unclear if this is the maximal radial extent of [\ion{C}{1}] 1--0 or the cutoff around 700~au is simply due to limited SNR. The depth of the central cavity is difficult to assess due to our angular resolution, i.e., the cavity size is comparable to the beam size. Overall, the radial morphology of [\ion{C}{1}] is similar to that seen in the majority of molecular species previously observed toward the IM~Lup disk \citep[e.g.,][]{Cleeves16_IMLup, Huang17, LawMAPSIII}. In Figure \ref{fig:radial_profiles}, we also show a radial profile of the peak intensity, computed using the full Planck function, which was extracted from a peak intensity map generated with the `quadratic' method of \texttt{bettermoments}. [\ion{C}{1}] 1--0 
shows an approximately constant brightness temperature of ${\approx}$10-15~K over the full extent of IM~Lup.

Next, we used the radial intensity profile and followed standard procedures \citep[e.g.,][]{Goldsmith99} to compute the [\ion{C}{1}] 1--0 optical depth and \ion{C}{1} column density (N$_{\rm{col}}$) profiles, which are shown in the middle and bottom panels, respectively, of Figure \ref{fig:radial_profiles}. All spectroscopic data were taken from the CDMS catalogue \citep{Muller01}. For simplicity, we assumed thermal line widths and adopted a constant excitation temperature of 50~K. This temperature is 10~K higher than the peak $^{12}$CO temperature measured in \citet{Law21} and is consistent with model expectations for the IM~Lup disk at the measured [\ion{C}{1}] 1--0 heights \citep[][also see Figure \ref{fig:zr_vs_models}]{Cleeves16_IMLup, pinte18}.

[\ion{C}{1}] 1--0 is optically thin over nearly all of the IM~Lup disk with the exception of the ring peak at ${\approx}$75~au, which shows an optical depth of order unity that rapidly declines beyond ${\approx}$100~au. Optically thin emission is consistent with the inferred [\ion{C}{1}] 1--0 brightness temperatures and thermochemical model predictions \citep{Kama16_TWHya_HD100546, Kama16_CI}. The column density profile is steep, with a peak N$_{\rm{col}}{\sim}$10$^{17}$~cm$^{-2}$ that declines by nearly four orders of magnitude in the outer disk. The IM~Lup disk has a break in its gas surface density at ${\approx}$400-500~au, which was inferred from CO observations \citep{Panic09, Zhang21}. The inferred \ion{C}{1} column density profile may show a similar break but at a smaller radius of ${\approx}$300~au, but higher spatial resolution data are required to confirm this.

For the sake of comparison with other disks, we also computed a disk-averaged N$_{\rm{col}}{\approx}$2$\times$10$^{16}$~cm$^{-2}$. Among disks with spatially-resolved \ion{C}{1}, this is approximately an order of magnitude lower than that of HD~142527 \citep{Temmink23} and HD~163296 \citep{Alarcon22} and a factor of a few less than the HD~169142 disk \citep{Booth23_HD169142}. Although IM~Lup and HD~142527 are both disks around T~Tauri stars, they show more than an order of magnitude difference in their \ion{C}{1} column densities. The IM~Lup disk is also an order of magnitude more massive \citep{Zhang21,Lodato23} than that of HD~142527 \citep{Temmink23}, which implies that the inferred \ion{C}{1} column density is not particularly sensitive to the total disk mass, as previously indicated in the models of \citet{Pascucci23}. The large mm dust cavity (${\approx}$100~au) of the HD~142527 disk leads to increased UV transparency and thus, also likely contributes to its larger \ion{C}{1} column densities.

\section{Discussion} \label{sec:discussion}

\subsection{\ion{C}{1} Emission Surface vs Model Predictions} \label{sec:model_predictions}

Here, we compare [\ion{C}{1}] 1--0 heights with the thermochemical model of the IM~Lup disk from \citet{Cleeves16_IMLup, Cleeves18} to assess the physical conditions at which \ion{C}{1} emits and across the \ion{C}{1}-to-CO transition. This model treats the gas and dust temperatures independently, which is critical in the upper disk layers where the two temperatures decouple. While this model does not explicitly predict the \ion{C}{1} abundance or emitting layer, \citet{Cleeves16_IMLup} infer a dense CO molecular layer, i.e., gas-phase CO abundance of ${\sim}$10$^{-4}$, that extends up to $z/r = 0.5$. This agrees well with our derived [\ion{C}{1}] 1--0 heights, as we expect atomic carbon to lie directly above this molecular layer.

\begin{figure}
\centering
\includegraphics[width=1\linewidth]{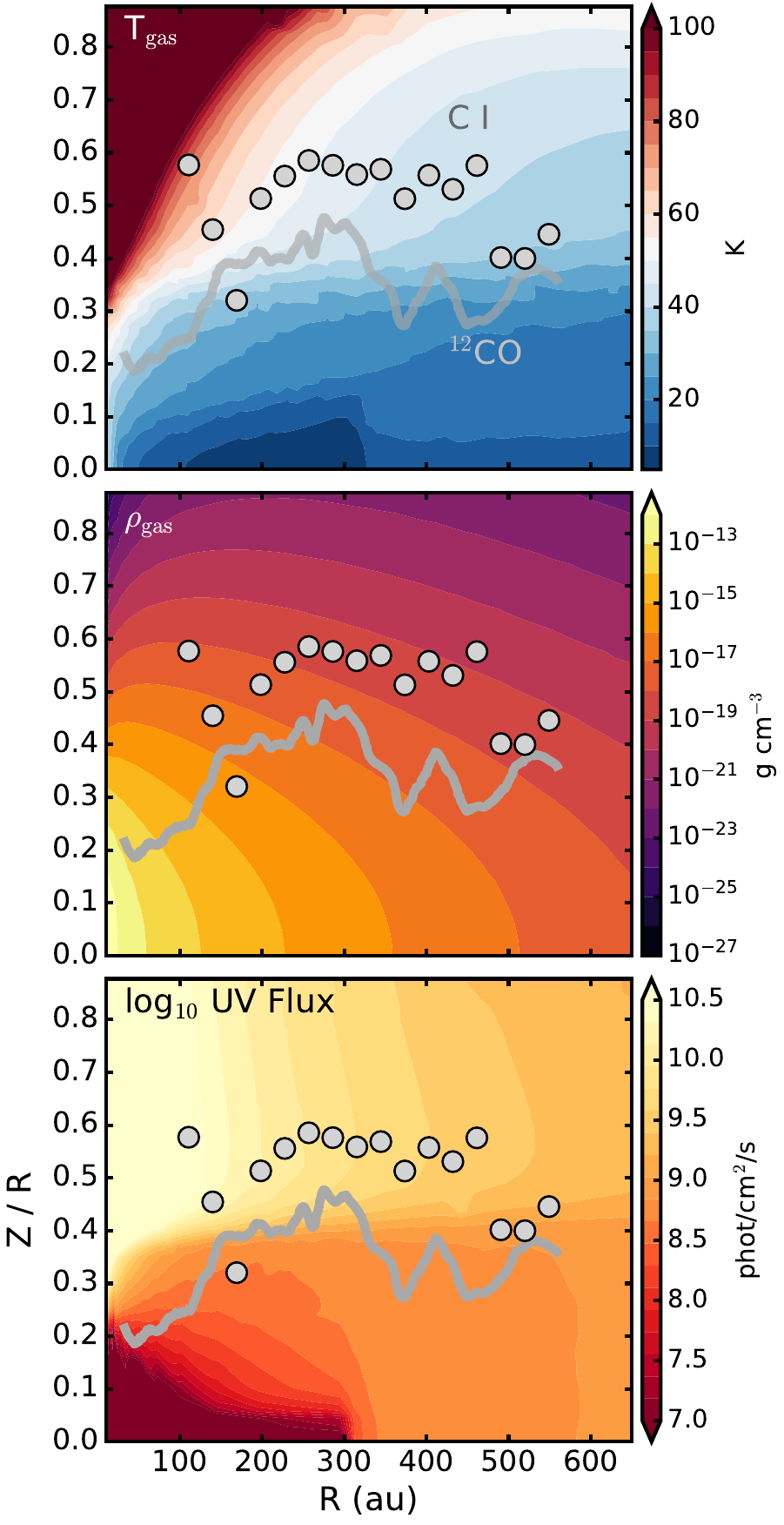}
\caption{[\ion{C}{1}] 1--0 (points) and $^{12}$CO J=2--1 (solid line) emitting heights compared to the thermochemical model (background) tuned to the IM~Lup disk from \citet{Cleeves16_IMLup} showing gas temperature (\textit{top}), gas density (\textit{middle}), and UV field (\textit{bottom}). The UV field is the stellar-integrated flux from 930-2000~\AA, including Ly$\alpha$, plus an external radiation field (G$_0=$4) from \citet{Cleeves16_IMLup}.}
\label{fig:zr_vs_models}
\end{figure}

The top two panels of Figure \ref{fig:zr_vs_models} show that the \ion{C}{1} emitting region corresponds to gas temperatures of ${\approx}$40--60~K and gas densities of 10$^{-18}$--10$^{-17}$~g~cm$^{-3}$. Given the large radial range over which [\ion{C}{1}] 1--0 is detected, these physical conditions are remarkably constant. The atomic-to-molecular transition, i.e., from \ion{C}{1} to the $^{12}$CO emitting surface, occurs in a thin vertical region between $z/r\approx$~0.4 to ${\lesssim}$0.5. Across this region, the temperature is typically no more than 10~K cooler, while gas densities are approximately only one order of magnitude lower. Thus, only modest changes in the gas physical conditions nonetheless correspond to substantial changes in the form of the gas present. The bottom panel of Figure \ref{fig:zr_vs_models} shows the UV field computed from the integrated stellar flux and external radiation field (G$_0$=4) from \cite{Cleeves16_IMLup}. The \ion{C}{1} emitting region corresponds to a UV flux of log$_{10}$ $(\rm{F}_{\rm{UV}} [\rm{phot} / \rm{cm}^2 / \rm{s}]) > 9.5$ and the \ion{C}{1}-to-CO transition has a lower bound of log$_{10}$ $(\rm{F}_{\rm{UV}} [\rm{phot} / \rm{cm}^2 / \rm{s}])\approx9$. The UV field in the \ion{C}{1} emitting region is dominated by the stellar contribution, which appears as a horizontal plateau in $Z/R$-space. At large radii (${>}$600~au), however, the external interstellar radiation field begins to dominate the total UV flux. It is at these radii where we no longer detect [\ion{C}{1}] 1--0, despite the IM~Lup disk having $^{12}$CO emission out to radius of ${\approx}$1000~au, which may suggest that the external UV field is photo-ionizing atomic carbon in the outer disk.

In addition, we can also compare the observed \ion{C}{1} heights to several thermochemical models tuned to other systems \citep[e.g.,][]{Jonkheid06, Kama16_TWHya_HD100546} as well as more generic disk models \citep[e.g.,][]{Jonkheid04, Kama16_CI, Ruaud22,Pascucci23}. In general, these models predict \ion{C}{1} emitting heights ranging from $z/r$~$\sim$~0.2--0.4 and typical $^{12}$CO heights of z/r~$\approx0.3$. Thus, while the $^{12}$CO emission surface approximately matches that measured in IM~Lup, the \ion{C}{1} heights are consistently under-predicted. Moreover, in many of these models, the \ion{C}{1} abundance remains high over a large range of $z/r$, while our observations of the IM~Lup disk show a considerably thinner \ion{C}{1} layer. This might suggest that the \ion{C}{1}-to-CO transition happens more quickly than these models predict, or alternatively, the IM~Lup disk is an especially atypical disk and the narrow, highly-elevated \ion{C}{1} emission may not be representative of other disks upon which these models were constructed. However, we note that the generic models of \citet{Jonkheid04} found that the C$^+$/\ion{C}{1}/CO transition occurs over a thin vertical region of $z/r\approx$0.4--0.6, which better matches our observed \ion{C}{1} heights.

\begin{figure*}
\centering
\includegraphics[width=\linewidth]{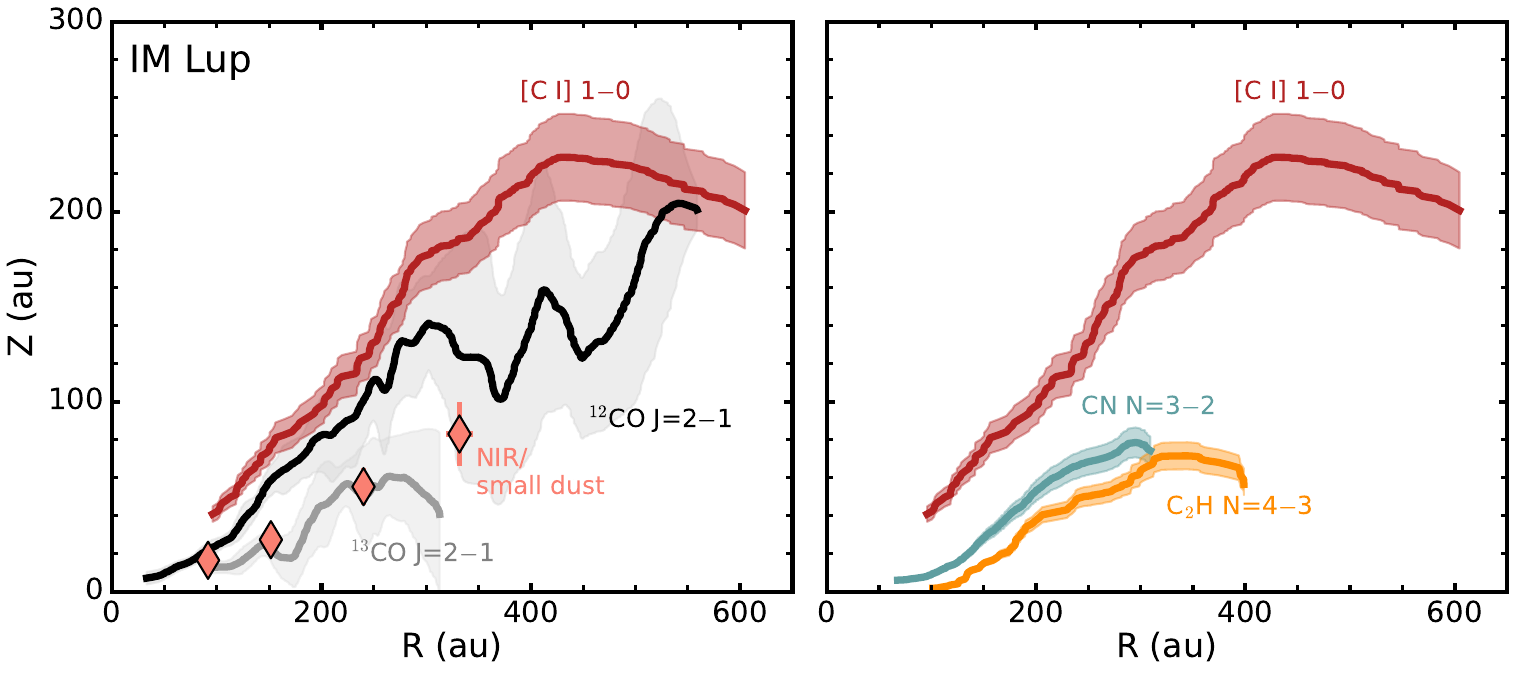}
\caption{[\ion{C}{1}] 1--0 emission surface in the IM~Lup disk versus that of the CO isotopologues and small dust heights (\textit{left}); and CN and C$_2$H (\textit{right}). The $^{12}$CO J=2--1 and $^{13}$CO J=2--1 heights are taken from \citet{Law21}, the NIR scattering surface is from \citet{Avenhaus18}, and the CN N=3--2 and C$_2$H N=4--3 heights are from \citet{Paneque23_subm}.}
\label{fig:surface_comparison}
\end{figure*}

\subsection{Vertical Structure of the IM~Lup Disk} \label{sec:IMLup_vert_str}

The left panel of Figure \ref{fig:surface_comparison} shows a comparison of the [\ion{C}{1}] 1--0 emission heights relative to $^{12}$CO J=2--1, $^{13}$CO J=2--1, and micron-sized dust. [\ion{C}{1}] 1--0 is clearly emitting from the uppermost disks layers and at all radii, is arising from a higher elevation than all other tracers. This matches expectations of the PDR-like structure of the disk upper layers, where the atomic carbon originates above the molecular layer due to the strong UV radiation which photodissociates $^{12}$CO. It is only when the UV field is sufficiently attenuated due to increased shielding can abundant molecular gas exist, as traced by $^{12}$CO. The relative ordering of $^{12}$CO and $^{13}$CO is instead set by the heights at which their optical depths become order unity, which in turn, reflects the relatively less abundant $^{13}$CO. Thus, the \ion{C}{1} emission surface probes a distinct chemical layer, i.e., optically-thin emission due to the changing UV field, rather than tracing a $\tau=1$ surface, which is the case for nearly all molecules, such as the CO isotopologues, whose vertical emission distribution has been mapped with ALMA.

\citet{Pascucci23} also recently argued that \ion{C}{1} is arising from the disk upper layers in the IM~Lup disk using the same archival ALMA datatset analyzed here by demonstrating that [\ion{C}{1}] 1--0 and $^{13}$CO J=2--1 show consistent disk-integrated spectral line profiles. As illustrated here, the direct extraction of surfaces from the image cubes provides a distinct advantage. By measuring the emitting height directly, we see that \ion{C}{1} emits from heights exceeding $^{13}$CO by 3$\times$. Thus, while spectra are useful tools in constraining the radial distribution of emission, it is challenging to extract vertical information without resolved observations. 

The right panel of Figure \ref{fig:surface_comparison} shows a comparison of the \ion{C}{1} heights with that of the CN N=3--2 and C$_2$H N=4--3 lines, which have emitting heights of $z/r\approx$~0.2-0.3 and $z/r~{\lesssim}$~0.2, respectively \citep{Paneque23_subm}. Although these molecules are thought to trace photochemistry and thus, be produced in disk regions with strong UV fields \citep[e.g.,][]{Bergin16, Cleeves18, Bosman21_MAPS7,Paneque22_CN}, both CN and C$_2$H are located below the \ion{C}{1} emitting layer. This suggests that the UV field strength necessary to produce abundant \ion{C}{1} lies above the warm molecular layer.

We do not identify any vertical substructure, i.e., a localized decrease in emitting heights, in the [\ion{C}{1}] 1--0 emitting surfaces, including at the location of substructures seen in the CO isotopologue emission surfaces \citep{Law21, Paneque_MAPS} or at the proposed location of the embedded giant planet identified via its $^{12}$CO velocity kinks \citep{Pinte20, Verrios22}. However, we emphasize the caveat that the current data are of only modest SNR and thus, we cannot definitively rule out the presence of vertical substructure. We lacked the sensitivity to detect subtle kinematic deviations \citep[e.g.,][]{Alarcon22} in the \ion{C}{1} Keplerian rotation profile, even if they are present. Likewise, no clear signatures of the photoevaporative wind proposed by \citet{Haworth17} were identified, at least within a radius of ${\approx}$600~au. To place robust constraints on the presence of spatially-localized \ion{C}{1} emission features, both higher angular resolution and more sensitive observations are needed.

\section{Conclusions} \label{sec:conlcusions}

Using high-angular-resolution ALMA archival data, we mapped the [\ion{C}{1}] 1--0 emission surface in the IM~Lup disk and provide direct observational confirmation that \ion{C}{1} is tracing the upper layers ($z/r$~$\gtrsim$~0.5) of a protoplanetary disk for the first time. \ion{C}{1} is tracing a narrow, optically-thin vertical slab and is located at scale heights almost double that of other disks with \ion{C}{1} heights inferred from unresolved observations or models. By using existing data of the $^{12}$CO and $^{13}$CO emitting heights, we spatially resolved the atomic-to-molecular gas transition zone. We did not find evidence for vertical substructures or spatially-localized deviations in \ion{C}{1} due, e.g., to either an embedded giant planet or a photoevaporative wind that have been previously proposed and confirm that [\ion{C}{1}] 1--0 emission is consistent with a Keplerian-rotating disk. We also computed a radially-resolved \ion{C}{1} column density profile and show that IM Lup has a disk-averaged \ion{C}{1} column density of 2$\times$10$^{16}$~cm$^{-2}$, which is $\approx$3-20$\times$ lower than that of other disks with spatially-resolved \ion{C}{1} observations.

The favorable geometry of IM~Lup, including its inclined and unusually large disk, allowed us to map out the \ion{C}{1} emission structure in detail with existing ALMA archival data of only modest SNR. While we can rule out the presence of large-scale \ion{C}{1} asymmetries, the current data quality precludes a detailed search for subtle, small-scale deviations. Sensitive, follow-up \ion{C}{1} observations of the IM~Lup disk present a unique opportunity to search for spatially-localized \ion{C}{1} asymmetries, which would provide a novel and perhaps unique way to study dynamical disk structures. In addition to further observations, dedicated, disk-specific modeling efforts of IM~Lup that incorporate the measured \ion{C}{1} heights would provide powerful constraints on the physical and chemical conditions of the upper disk layers. In particular, the IM~Lup disk is an ideal candidate for such efforts given its unique set of UV-sensitive atomic and molecular tracers (e.g., \ion{C}{1}, CN, C$_2$H) as well as NIR/scattered light images of its small dust distribution that sets the UV opacity, which, taken together, will allow for a detailed mapping of its radial and vertical UV field. \\ 


The authors thank the anonymous referee for valuable comments that improved the content of this work. This paper makes use of the following ALMA data: ADS/JAO.ALMA\#2015.1.01137.S and 2018.1.01055.L. ALMA is a partnership of ESO (representing its member states), NSF (USA) and NINS (Japan), together with NRC (Canada), MOST and ASIAA (Taiwan), and KASI (Republic of Korea), in cooperation with the Republic of Chile. The Joint ALMA Observatory is operated by ESO, AUI/NRAO and NAOJ. The National Radio Astronomy Observatory is a facility of the National Science Foundation operated under cooperative agreement by Associated Universities, Inc. Support for C.J.L. was provided by NASA through the NASA Hubble Fellowship grant No. HST-HF2-51535.001-A awarded by the Space Telescope Science Institute, which is operated by the Association of Universities for Research in Astronomy, Inc., for NASA, under contract NAS5-26555.

%\vspace{5mm}
\facilities{ALMA}


\software{\texttt{alfahor} \citep{Paneque_MAPS}, Astropy \citep{astropy_2013,astropy_2018}, \texttt{bettermoments} \citep{Teague18_bettermoments}, CASA \citep{McMullin_etal_2007}, \texttt{cmasher} \citep{vanderVelden20}, \texttt{emcee} \citep{Foreman_Mackey13}, \texttt{GoFish} \citep{Teague19JOSS}, \texttt{keplerian\_mask} \citep{rich_teague_2020_4321137}, Matplotlib \citep{Hunter07}, NumPy \citep{vanderWalt_etal_2011}}



\clearpage
\appendix

\section{[\ion{C}{1}] 1--0 Channel Maps} \label{sec:appendix_channel_maps}

\begin{figure*}[!h]
\centering
\includegraphics[width=\linewidth]{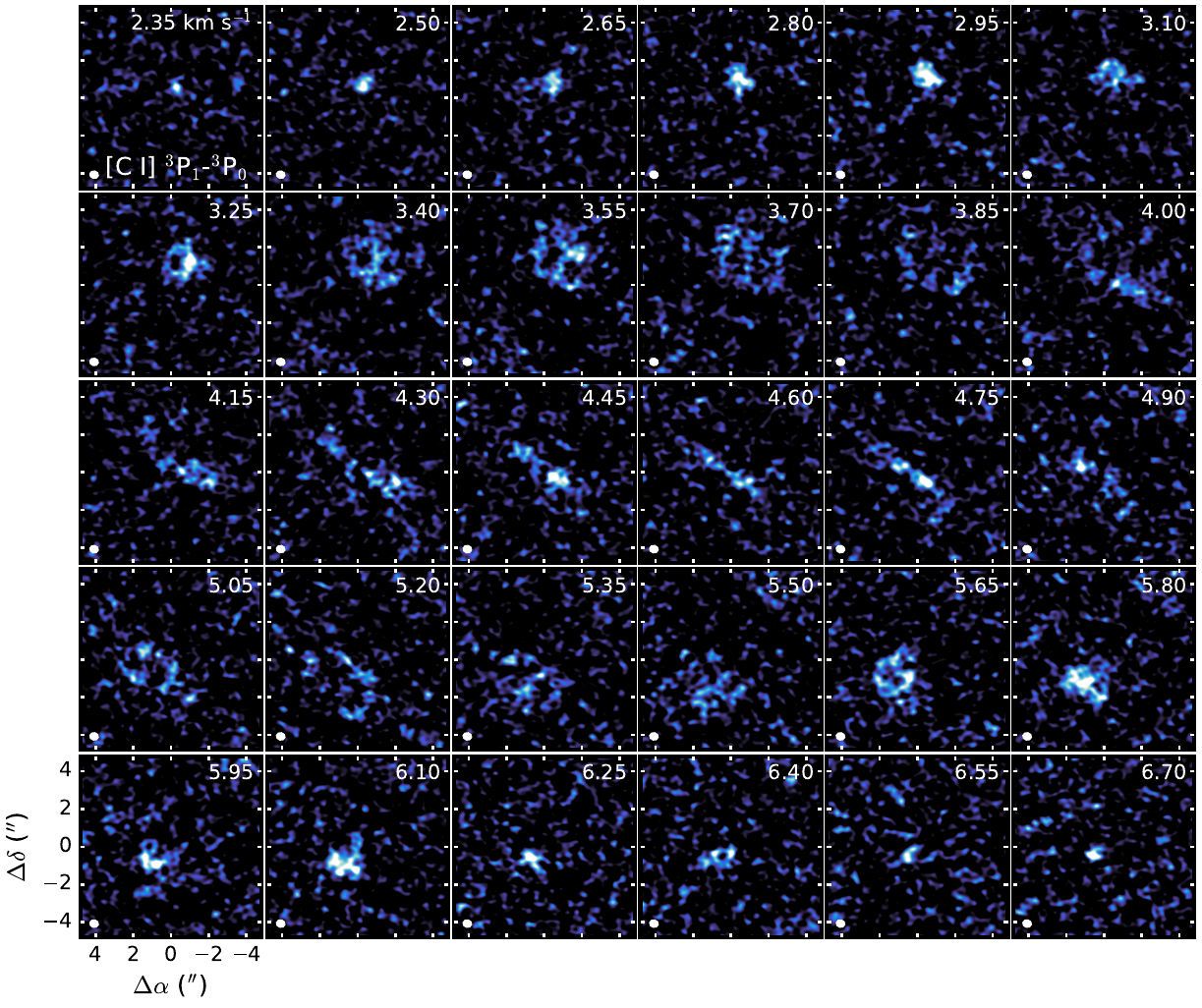}
\caption{Channel maps of [\ion{C}{1}] 1--0 emission in the IM~Lup disk. The synthesized beam is shown in the lower left corner of each panel and the LSRK velocity in km~s$^{-1}$ is printed in the upper right.} 
\label{fig:app_chans}
\end{figure*}

Figure \ref{fig:app_chans} shows a complete set of [\ion{C}{1}] 1--0 channel maps in the IM~Lup disk.

\clearpage


\bibliography{CI_IMLup}{}
\bibliographystyle{aasjournal}



\end{document}